\documentclass[preprint,12pt]{elsarticle}

\usepackage{graphicx}
\usepackage{amssymb}
\usepackage{array}
\usepackage{booktabs}
\usepackage{tabu}
\usepackage{dcolumn}
\usepackage{amsmath}
\usepackage{amsfonts}
\usepackage{subfigure}
\usepackage{graphicx}
\usepackage{dcolumn}
\usepackage{bm}
\usepackage{caption}
\usepackage{lineno}

\begin{document}

\begin{frontmatter}


\title{Inflationary universe in the presence of  a minimal measurable length}


\author{A. Mohammadi$^{\P}$\footnote{Electronic address~:~\upshape{abolhassanm@gmail.com}}}
\author{Ahmed Farag Ali$^{\ddagger \S}$\footnote{Electronic address~:~\upshape{ahmed.ali@fsc.bu.edu.eg,~~afali@fsu.edu }}}
\author{ T. Golanbari $^{\P}$}
\author{A. Aghamohammadi$^{\pounds}$}
\author{Kh. Saaidi $^{\P}$}
\author{Mir Faizal$^{\sharp}$}

\address{$^{\P}$Department of Physics, Faculty of Science, University of Kurdistan, Pasdaran Street, P.O. Box 66177-15175 Sanandaj, Iran,\\
$^{\ddagger}$Department of Physics, Florida State University, Tallahassee, FL 32306, USA,
$^{\S}$Department of Physics, Faculty of Science, Benha University, Benha, 13518, Egypt,
$^{\pounds}$Sanandaj Branch, Islamic Azad University, Pasdaran Street, P.O. Box 618 Sanandaj, Iran,
$^{\sharp}$Department of Physics and Astronomy, University of Waterloo, Waterloo, Ontario, N2L 3G1, Canada.}

\begin{abstract}
In this paper, we will study the effect of having a minimum measurable length on inflationary cosmology.
We will analyze the inflationary cosmology in the Jacobson approach. In this approach, gravity is viewed
as an emergent thermodynamical phenomena. We will demonstrate that the existence of a minimum
measurable length will  modify the Friedmann equations in the Jacobson approach. We will use this  modified
Friedmann equation to analyze the effect of minimum measurable length scale on inflationary cosmology.
This analysis will be performed using the Hamiltonian-Jacobi approach. We compare our results to recent data, and find that our model may agree with the recent data.
\end{abstract}



\end{frontmatter}


\section{Introduction}
It is known that   a connection exists between the thermodynamics and gravity.
This connection was first investigated in the works of Bardeen, Carter and Hawking \cite{Bardeen}.
 In this work, it was suggested that an analogy exists between the laws of thermodynamics and
 gravitational physics. However, the existence of this connection was only established after the discovery
 of Hawking radiation  \cite{Hawking}. The Hawking radiation comes from black holes, as black holes behave as
 hot bodies with a temperature proportional to the surface gravity of the black hole.
 The black holes also have an entropy associated with them, and this entropy is proportional to the area of the
 horizon \cite{Bekenstein,Amelino-Camelia01,Banerjee01,Garattini01}. Thus, the black hole physics establishes a connection between the laws of thermodynamics
 and gravitational physics.
Motivated by the relation between thermodynamics and gravity, it has been proposed that gravity is actually
only an emergent thermodynamic phenomena. This formalism in which gravity is described
as a emergent phenomena is called the Jacobson formalism \cite{Jacobson}. In this formalism,
the Einstein  field equation can be derived from the first law of thermodynamics $dQ=T dS$  \cite{Cai,Akbar:2006kj}.

It may be noted that  the Hawking radiation is connected to  the uncertainty principle in quantum mechanics
\cite{Eliasentropy,Adler,Cavaglia:2003qk,Cavaglia:2003qk1,Majumder,Majumder1,Majumder2}. This connection is established by modeling the
black hole as a $n$-dimensional sphere whose  radius is related to the Schwarzschild radius.
The emitted particles from the black hole obey the uncertainty principle, as
the Hawking radiation is a  purely quantum mechanical process.
This fact can be used to derive the thermodynamical properties of the black hole  \cite{Adler,Cavaglia:2003qk,Cavaglia:2003qk1}.
It is known that the temperature of a black hole increases as the size of the black hole decreases.
This temperature tends to infinity as the mass size of the black hole approaches zero, and this in turn leads
to a catastrophic evaporation of the black hole.

There are strong indications that there exists a minimum measurable length
scale for the spacetime \cite{guppapers,BHGUP,BHGUP1,BHGUP2,BHGUP3,BHGUP4,BHGUP5,BHGUP6,Scardigli,kmm,kempf,brau}.
In fact, the existence of a minimum measurable  length scale is universal
feature of almost all approaches to quantum gravity \cite{guppapers,BHGUP,BHGUP1,BHGUP2,BHGUP3,BHGUP4,BHGUP5,BHGUP6,Scardigli}.
The existence of the minimum measurable length scale is not consistent with the usual uncertainty principle.
This is because according to the usual uncertainty principle, it is possible to measure the length to an arbitrary
accuracy, if the momentum is not measured. However, it is possible to generalize
the usual uncertainty principal  to a generalized uncertainty principle (GUP),
such that this new uncertainty principle is consistent with the existence of a minimum measurable length.
As the Hawking radiation depends on the uncertainty principle, the modification of the usual
uncertainty principle to the generalized uncertainty principle, also modifies the
thermodynamics of the black holes  \cite{Eliasentropy,Adler,Cavaglia:2003qk,Cavaglia:2003qk1,Majumder,Majumder1,Majumder2}.
 In fact, the usual relation between the entropy
and the area of a black hole gets significantly modified due to the GUP
 \cite{Eliasentropy,Majumder,Majumder1,Majumder2} .

As the Friedmann equations are viewed as thermodynamical relations in the Jacobson formalism, the modification
of the thermodynamics by the GUP will also modify the Friedmann equations in the Jacobson formalism. Such
modification to the Friedmann equations  has been recently studied \cite{Awad:2014bta}. It has also been
  demonstrated that the maximum energy density and a general nonsingular evolution is
independent of the equation of state and the spacial curvature $k$. This state of maximum energy density
was reached in a finite time  \cite{Adel}. So, in this model, the big big bang singularity is not accessible, and the  energy density in the
spacetime  cannot be extended beyond Planck density. This bound on the energy
density of spacetime  occurs because  of the GUP.  As the Friedmann equations get modified due to the GUP,
we expect the inflationary cosmology also to get modified due to the GUP in the thermodynamical approach.
So, in this work we will study the inflationary cosmology using the Jacobian formalism. We will also study the
effect of GUP deformed thermodynamics on the inflationary cosmology

It is possible to study the inflationary cosmology by using a potential
to derive the properties of the inflationary cosmology.
Thus, a specific form of the potential is chosen, and the model of the inflationary
cosmology depends on the details
of the potential chosen \cite{1,1a,1b,1c}.  However, in addition to all of other possible ways of considering inflation \cite{Amelino-Camelia22,Garattini}, it is also possible to use the
  Hamilton-Jacobi formalism to model the inflationary universe \cite{2,2a,2b,2c,2d}.
  In this  formalism,    the Hubble parameter is expressed in term of a scalar field.
 It is possible to
  deducing a form of the potential in the the Hamilton-Jacobi formalism.
  Furthermore, it is also possible to obtain an exact solution for the scalar
field in this approach. Thus, we will use the Hamilton-Jacobi formalism in this work.

\section{Inflation and Minimal Length}\label{RayChud}
\label{inflationGUP}

In this section, we will first   review the modification to the Friedman equations in the
  thermodynamics relation  approach  \cite{Jacobson,Cai,Akbar:2006kj,Hayward,Awad:2014bta}.
The GUP \cite{guppapers,BHGUP,BHGUP1,BHGUP2,BHGUP3,BHGUP4,BHGUP5,BHGUP6} is known to modify the area-entropy law
\cite{Eliasentropy,Majumder,Majumder1,Majumder2,Adler,Cavaglia:2003qk,Cavaglia:2003qk1}. The existence of a minimum
measurable length originates as an intriguing prediction of various frameworks of
quantum gravity such as string theory \cite{guppapers} and black hole
physics \cite{BHGUP,BHGUP1,BHGUP2,BHGUP3,BHGUP4,BHGUP5,BHGUP6}. This implies a direct modification of the
standard uncertainty principle \cite{guppapers,BHGUP,BHGUP1,BHGUP2,BHGUP3,BHGUP4,BHGUP5,BHGUP6,Scardigli,kmm,kempf,brau},
$\Delta x \geq
 \frac{\hbar}{\Delta p}\left[1+ \frac{\beta~
\ell_{P}^2}{\hbar^2} (\Delta p)^2\right], \label{GUP}
$,
where
$\ell_{P}$ is the Planck length and $\beta$ is a dimensionless
constant which is generated from some quantum gravitational effect.   This modification of the thermodynamics
 of a black hole also modifies the   Bekenstein-Hawking
 entropy of a  black hole
\cite{Eliasentropy,Majumder,Majumder1,Majumder2,Adler,Cavaglia:2003qk,Cavaglia:2003qk1,Awad:2014bta})
\begin{eqnarray} \frac{dS}{dA}= \frac{\alpha}{8 \ell_P^2} \frac{1}{A \left[1-
\sqrt{1-\frac{\alpha}{A}}~\right]}. \label{exact}
\end{eqnarray}
where $\alpha= 4 \beta \ell_P^2 \pi$. It is possible to calculate the exact form of the
GUP deformed Friedmann equations in the thermodynamic approach   \cite{Awad:2014bta},
\begin{equation}
\frac{8 \pi G}{3} (\rho-\Lambda) = \frac{1}{2} \left(H^2+\frac{k}{a^2}\right) +{4\,\pi \over 3 \alpha}\, \left[1-\mathcal{F}^{\frac{3}{2}}(H)\right], \label{MFRA}
\end{equation}
\begin{equation}
- 4 \pi G(\rho+p)= \frac{\frac{\alpha}{8\, \pi}\left(H^2+\frac{k}{a^2}\right)\left(\dot{H}-\frac{k}{a^2}\right)}
{\left[1-\mathcal{F}^{\frac{1}{2}}(H)\right]}.\label{MFRB}
\end{equation}
in which the function $\mathcal{F}(H)$ is defined by
\begin{equation}\nonumber
\mathcal{F}(H) = 1-\frac{\alpha}{4\, \pi}\,\left(H^2+\frac{k}{a^2}\right).
\end{equation}

Now we can analyze the inflationary cosmology in the presence of a minimum measurable length.
In this regards, the Hamilton-Jacobi approach is picked out as an appropriate formalism for studying
inflation and its interesting prediction, quantum perturbations.
We set $k=0$ in our case for studying inflation\footnote{ Inflation area is a epoch of the universe, where
it undergoes an extremely accelerated expansion. In other word, the scale factor of the universe grows fast during the inflationary times.
Besides, we are considering a slow-rolling inflation where the scalar field moves slowly, as a ball in liquid. Consequently, the energy density
of scalar field remains almost constant and the universe continues expansion. Because of rapid grows of universe and presence of slow moving scalar
field, the constant parameter $k$, which appears as $k/a^2$ in the Friedmann equation, could be ignored \cite{linde}.}.
Considering the density and the pressure given by
\begin{equation}
\rho= \frac{1}{2} \dot{\phi}^2+V(\phi), \qquad p= \frac{1}{2} \dot{\phi}^2-V(\phi)
\end{equation}
the modified Friedman equation will be as follows:
\begin{equation}\label{Fri}
 \frac{1}{2} H^2 + \frac{4\pi}{\alpha} \left[1 - \mathcal{F}^{3/2}\right] = \frac{8 \pi G}{3}
 \left(\frac{\dot{\phi}^2}{2} + V (\phi)\right)
\end{equation}
Note that, from the Friedmann equation (\ref{Fri}), it is realized that the expression $\alpha H^2 / 4\pi$ must be always smaller than unity. In the rest of the work, we take into account this and expand the term up to the third order.\\
The Hamilton-Jacobi formalism is utilized as a strong tool to provide numerous inflationary models with exactly
known analytic solutions for the background expansion \cite{Bauman}. In this formalism, instead of potential, the Hubble parameter is introduced as a function of scalar field $H:=H(\phi)$. Consequently, the time derivative of the Hubble parameter is rewritten as $\dot{H}=\dot\phi H'$, where prime denotes derivative with respect to scalar field. From the second Friedmann equation (\ref{MFRB}), we have the time derivative of the scalar field as
\begin{equation}
\dot\phi = -{1 \over 4\pi G}\; {H' \over \left( 1 + {\alpha \over 16\pi}\;H^2 \right)}.
\end{equation}
Substituting the relation in the first Friedmann equation comes to the Hamilton-Jacobi equation

\begin{eqnarray}\label{H-J}
\frac{1}{2} H^2 & + & \frac{4\pi}{\alpha} \left[1 - \mathcal{F}^{3/2}(H)\right] \nonumber \\
  &  & \hspace{1cm} = \frac{8 \pi G}{3}
\left(\frac{1}{2(4 \pi G)^2} \frac{{H'}^2}{ (1 + \frac{\alpha}{16\pi}H)^2 }+  V (\phi)\right)
\end{eqnarray}
where the predicted potential of the model could easily be estimated, and the general behavior of the potential is investigated easily. \\

\subsection{Perturbation}
The most interesting aspect of the inflationary scenario is that the scenario predicts quantum perturbation in the very early times of the universe evolution. There are three types of perturbation known as scalar, vector, and tensor perturbations, and the most important ones are scalar and tensor perturbations. Scalar fluctuations become seeds for cosmic microwave background (CMB)
anisotropies or for large scale structure (LSS) formation. Therefore, by measuring the spectra of the CMB anisotropies
and density distribution, the corresponding primordial perturbations could be determined. We briefly explain the situation.
Involving only the scalar perturbation in metric, one has
\begin{equation}\label{metricS}
ds^2 = a^2(\tau) \Big[ (1+A)d\tau^2 - 2\partial_i B d\tau dx^i -
\big(\delta_{ij} + 2C\delta_{ij} + 2(\partial_i\partial_j E - {1 \over 3} \delta_{ij}\nabla^2 E) \big)dx^i dx^j \Big]
\end{equation}
and the action for canonical scalar field is stated as
\begin{equation}\label{actionS}
S = \int d\tau d^3x \Big( R + {1 \over 2} g^{\mu\nu} \partial_\mu \partial_\nu - V(\phi)  \Big).
\end{equation}
Mukhanov-Sasaki equation plays an important role to debate scalar perturbation of inflation that is derived by variation of action (the second order of action in fluctuation) . In order to consider the dynamical equation in linear order, the action is required at quadratic order in fluctuations. Applying the spatially-flat gauge (where $C=E=0$), it is realized that fluctuations in the geometric part of the action are appeared by a factor of slow-rolling parameter $\epsilon$(that is very small). Then, the fluctuation in the geometry could be disregarded and the intended equation could be derived only by perturbing the inflaton part. Finally, the Mukhanov-Sasaki equation could be found by variation of obtained action. Since the geometric part could be ignored in the mentioned gauge and the scalar perturbation only appears at the scalar field part of action.
\begin{equation}\label{actionSperturb}
S_2 = {1 \over 2} \int d\tau d^3x \left(  (\upsilon')^2 - (\nabla \upsilon)^2 -  \Big[ {a'' \over a} - a^2 V_{,\phi\phi} \Big] \upsilon^2  \right).
\end{equation}
where $\upsilon=a\delta\phi$ ($\phi = \bar{\phi} + \delta\phi$ and $\bar{\Box}$ indicates the background parameter). Generally, if someone does not ignore the metric fluctuation, finally they arrives at the same equation for the amplitude of scalar perturbation as one arrive for the above case at horizon crossing that is our interest point \cite{bauman11}. \\
It could be found out that the calculation for perturbation in our case is not modified. However, it should be noted that the evolution of the Hubble parameter and scalar field, which appear in the amplitude of scalar perturbation and determine the dynamics of parameters, are certainly modified.  \\
The above explanation gives a qualitative description of deriving the amplitude of scalar perturbation (for more detail refer to \cite{bauman11,bauman111}). However, if one is interested in qualitative description it could be found in \cite{Campo}. \\
Employing the amplitude of scalar perturbations from \cite{Campo}, the scalar spectra index is read as \cite{Riotto}

\begin{equation}\label{ns}
n_s - 1 = { d\ln( \mathcal{P}_s ) \over d\ln(k)} = 2\eta_H - 4\epsilon_H,
\end{equation}
where $\epsilon_H$ and $\eta_H$ are the first and second slow-rolling parameters, given by \cite{Riotto, Weinberg}

\begin{equation}\label{SRP}
\epsilon_H = -{\dot{H} \over H^2}; \qquad \eta_H = -{|\ddot\phi| \over H|\dot\phi|}.
\end{equation}

Besides scalar fluctuations, the inflationary scenario predicts tensor fluctuations, which is known as a gravitational
wave, too. The produced tensor fluctuations induce a curved polarization in the CMB radiation and increase the
overall amplitude of their anisotropies at a large scale. The physics of the early Universe could be specified by fitting
the analytical results the of CMB and density spectra to corresponding observational data. In this case, the perturbed metric could be written as
\begin{equation*}
ds^2 = a^2(\tau) \Big[ d\tau^2 - (\delta_{ij}+2E_{ij})dx^i dx^j \Big]
\end{equation*}
On the other hand, energy-momentum tensor has no contribution in the tensor perturbation. Then, only the geometric part of the action should be considered where it should be computed at quadratic order in fluctuations. For a Einstein-Hilbert action, the perturbation of the action at quadratic order is derived as
\begin{equation}\label{actionT}
S_2 = {M_p^2 \over 8} \int d\tau d^3x \Big( (E'_{ij})^2 - (\nabla E_{ij})^2 \Big)
\end{equation}
Finally using almost the same method as for scalar perturbation, the amplitude of tensor perturbation is obtained, as $\mathcal{P}_T = {16\pi \over m_p^2}\left({H \over 2\pi}\right)^2$. After doing some algebraic analysis, and obtaining the amplitude of tensor perturbations \cite{Campo}, the tensor spectra index is derived as \cite{Riotto}

\begin{equation}
n_T =  { d\ln( \mathcal{P}_T ) \over d\ln(k)} = -2\epsilon_H.
\end{equation}

The imprint of tensor fluctuations on the CMB bring this idea to indirectly determine its contribution to power
spectra by measuring CMB polarization \cite{Terrero}. Such a contribution could be expressed by the $r$ quantity, which is
known as tensor-to-scalar ratio and represents the relative amplitude of tensor-to-scalar fluctuations, $r = \mathcal{P}_T / \mathcal{P}_s$. Therefore, constraining $r$ is one of the main goals of the modern CMB survey, and is given by

\begin{equation}\label{r}
r = { \mathcal{P}_T \over \mathcal{P}_s } = {1 \over 2\pi G}\; {H'^2 \over H^2 \left( 1 + {\alpha \over 16\pi}\;H^2 \right)^2} .
\end{equation}

Inflation lasts until the slow roll parameter $\epsilon_H$ approaches unity. Then, the final value of scalar field could be read from Eq.(\ref{SRP}). In order to estimate the field value in the beginning of inflation, the common approach is to employ the number of e-folds equation. The number of e-folds, indicated by $N$, is expressed as following

\begin{equation}\label{efold}
N = \int_{\phi_i}^{\phi_e} {H \over \dot\phi} \; d\phi,
\end{equation}
where the subscripts "$i$" and "$e$" respectively denote the initial and end of inflation. Integrating the equation for specific Hubble parameter, and using $\epsilon_H=1$, the initial value of scalar field could be calculated, which will be done in the following subsection. \\

\subsection{Potential}
The general form of the potential was introduced in Eq.(\ref{H-J}). In order to study the
behavior of the potential during the inflationary times, the potential could be drew in term of scalar field. Rewriting the Hamilton-Jacobi equation (\ref{H-J}), the potential could be expressed by

\begin{eqnarray}
 V(\phi) &=& \frac{3}{ 8 \pi G} \left[ \frac{H^2}{2} - \frac{4 \phi}{\alpha}\Big(1 - \mathcal{F}^{3/2}(H)\Big)\right] \nonumber \\
  & & \hspace{2cm} - \frac{1}{2 (4 \pi G)^2}\frac{{H'}^2}{(1 + \frac{\alpha}{16 \pi} H^2)^2}
 \end{eqnarray}
This is a the general form of the potential in term of the Hubble parameter. To investigate the behavior of the potential, we need to introduce a function for the Hubble parameter in term of scalar field, that is what we are going to do in next subsection.\\

\subsection{Attractor Behavior}
One of the absorbing aspect of the Hamilton-Jacobi formalism is that it allows to simply consider the attractor behavior of the model. The common approach is to suppose a homogeneous perturbation for Hubble parameter as $H = H_0 + \delta H$ \cite{Lyth},
and insert it in the Hamiltonian-Jacobi equation up to the first order. If the expression $\delta H(\phi) / H_0(\phi)$ approaches to zero with increasing time, the attractor condition could be satisfied. In the following lines, the calculation is explained step by step for more clarity. The Friedmann equation is expressed by
\begin{equation*}
 \frac{1}{2} H^2 + \frac{4\pi}{\alpha} \left[1 - \Big( 1 - {\alpha \over 4\pi}H^2 \Big)^{3/2}\right] = \frac{8 \pi G}{3}
 \left(\frac{1}{2(4 \pi G)^2} \frac{{H'}^2}{ (1 + \frac{\alpha}{16\pi} H)^2 } + V (\phi)\right)
\end{equation*}
Now, we impose the perturbation, $H \rightarrow H_0+\delta H$, to the above equation. In the following lines, we consider it for every term one by one
\begin{eqnarray*}
\bullet & \quad &  H^2 = (H_0 + \delta H)^2 = H_0^2 (1 + \delta H / H_0)^2 = H_0^2 (1 + 2\delta H/ H_0) = H_0^2 + 2H_0 \delta H \hspace{3.8cm}
\end{eqnarray*}
\begin{eqnarray*}
\bullet & \quad &  1 - \Big( 1 - {\alpha \over 4\pi}H^2 \Big)^{3/2} =  1 - \Big( 1 - x \Big)^{3/2} \simeq 1 - \Big( 1- {3 \over 2} x + {3 \over 8} x^2 \Big) = {3 \over 2} {\alpha \over 4\pi} H^2 - {3 \over 8} \Big({\alpha \over 4\pi}\Big)^2 H^4 \hspace{2cm}\\
  &  & \hspace{6cm} ={3 \over 2} {\alpha \over 4\pi} \big( H_0^2 + 2H_0 \delta H \big) - {3 \over 8} \Big({\alpha \over 4\pi}\Big)^2 \big( H_0^4 + 4H_0^3 \delta H \big)  \\
  &  & \hspace{6cm} = {3 \over 2} {\alpha \over 4\pi} H_0^2 \Big( 1 - {\alpha \over 16\pi } H_0^2 \Big)
                        + 3 {\alpha \over 4\pi} H_0 \Big( 1 - {\alpha \over 8\pi } H_0^2 \Big)\delta H
\end{eqnarray*}
in above the dimensionless quantity $x$ is defined as $x=\alpha H^2 / 4\pi$.
\begin{eqnarray*}
\bullet & \quad &  {H' \over (1 + x/4)^2} = {H' \over (1 + x/2 + x^2/16)} \simeq {H_0'^2 \Big( 1 + 2{\delta H' \over H'_0} \Big) \over 1 + {\alpha \over 8\pi}H_0^2\Big( 1 + 2{\delta H \over H_0} \Big) + {1 \over 16}\Big({\alpha \over 4\pi}\Big)^2H_0^4\Big( 1 + 4{\delta H \over H_0} \Big) } \hspace{2.8cm}  \\
  &  & \hspace{5cm}  ={ H_0'^2 + 2 H_0' \delta H' \over  \Big( 1 + {\alpha \over 8\pi} H_0^2 + {1 \over 16}\big({\alpha \over 4\pi}\big)^2 H_0^4 \Big) +  \Big(  {\alpha \over 4\pi} H_0 + {1 \over 4}\big({\alpha \over 4\pi}\big)^2 H_0^3 \Big)\delta H } \\
  &  & \hspace{5cm}  ={ H_0'^2 + 2 H_0' \delta H' \over A_1 + A_2 \delta H } =  { H_0'^2  \over A_1 + A_2 \delta H } + { 2 H_0' \delta H' \over A_1 + A_2 \delta H } \\
  &   &  \hspace{5cm} = {H_0' \over A_1}\Big( 1 - {A_2 \over A_1} \delta H \Big) + {H_0'\delta H' \over A_1}\Big( 1 - {A_2 \over A_1} \delta H \Big)
\end{eqnarray*}
where
\begin{eqnarray}
 A_1&=& 1 + \frac{\alpha}{8\pi}H_0^2 + \frac{1}{16}\frac{\alpha^2}{(4\pi)^2} H_0^4
 \nonumber \\
 A_2 &=& \left(\frac{\alpha}{4 \pi}H_0 + \frac{1}{4} \frac{\alpha^2}{(4 \pi)^2}H_0^3 \right)
\end{eqnarray}
So far, we explain every term of the Friedmann equation separately. Now, by substituting them in the Friedmann equation and keeping only the first order terms, one arrives at
\begin{equation*}
H_0\delta H + \Big( {\alpha \over 4\pi} \Big) \left[ 3\Big( {\alpha \over 4\pi} \Big) H_0 - {3 \over 2} \Big( {\alpha \over 4\pi} \Big)^2 H_0^3 \right] \delta H = {-1 \over 12\pi G} \left[ {{H'}^2_0 A_2 \over A_1^2}\delta H + {H_0' \over A_1} \delta H' \right]
\end{equation*}
and finally, by rearranging the above equation, we could derive the equation of the main manuscript, namely as
\begin{equation*}
\left[ H_0 + 3 H_0 - {3 \over 2} \Big( {\alpha \over 4\pi} \Big) H_0^3 + {1 \over 12\pi G} {H_0' A_2 \over A_1^2}  \right] \delta H = {-1 \over 12\pi G}  {H_0' \over A_1} \delta H'
\end{equation*}
and finally
\begin{equation*}
 {\delta H' \over \delta H}  = {-12\pi G A_1 \over H_0'}\left[ 4 H_0 - {3 \alpha \over 8\pi}  H_0^3 + {1 \over 12\pi G} {H_0' A_2 \over A_1^2}  \right] = \mathbb{H}_0(\phi)
\end{equation*}
Now we have
\begin{equation}\label{attractor}
 \delta  H(\phi) = \delta H_0 \exp \int \mathbb{H} (\phi) d\phi.
\end{equation}
A specific function of scalar field for the Hubble parameter is required to exactly find out that the model could have the attractor behavior. It is postponed to the next section where a function for the Hubble parameter is proposed.\\

\section{Typical Example}
\begin{figure}[h]
\centering
\subfigure[ $V(\phi)-\phi$ ]{\includegraphics[width=4.5cm]{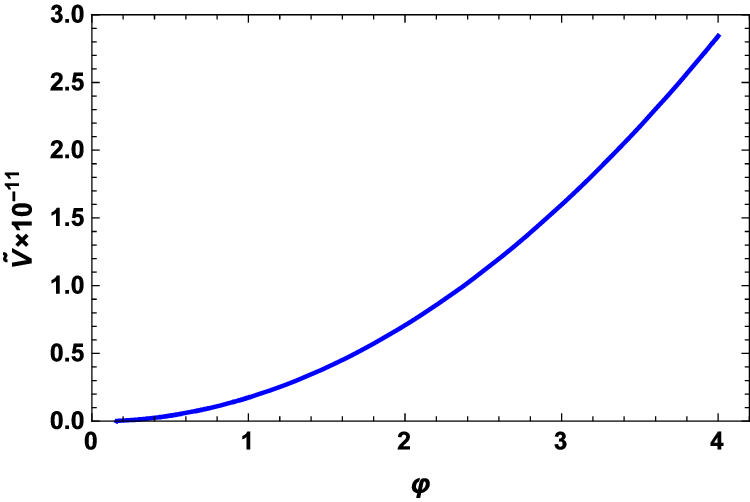}\label{F02}}
\hspace*{5mm}
\subfigure[ $\omega -\phi$]{\includegraphics[width=4.5cm]{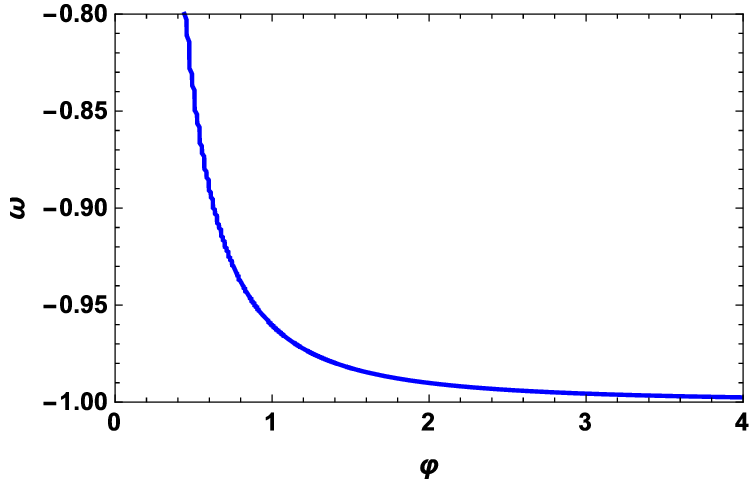}\label{F03}}
\caption{{\footnotesize a) the potential b) the equation of state parameter : have been plotted versus scalar field for $\beta=9\times10^{-4}$, $\beta \mathcal{H}_1^2=1\times10^{-14}$ and $N=60$. The parameter $\tilde{V}$ is defined as $V/M_p^4$, where $M_p$ is Planck mass $M_p^2=G^{-1}$. Also $\varphi$ is the dimensionless scalar field $\varphi \equiv \phi/M_p$. }} \label{potomega}
\end{figure}

To go further, and investigate the result in more clear detail, it is necessary to propose a specific function for the Hubble parameter in term of scalar field. In this regards, we assume that there is $H=\mathcal{H}_1 \phi$, in which $\mathcal{H}_1$ is a constant. Substituting this definition of $H(\phi)$, the situation will be studied in more detail.\\
Given $H(\phi)$, the the initial features, namely the potential and equation of state parameter, of the model could be considered at the first step. The parameters have been depicted in Fig.\ref{potomega} that shows a desirable situation: the potential decreases by passing time and is smaller than Planck energy density. At the beginning of inflation, the scalar field (inflaton) stands on top of the potential, and it rolls down to the minimum by increasing time, or with reducing scalar field. Then, it is realized that the inflation in this work could be classified as a member of \textit{"Large Field Model"} class of inflation so that the scalar field interval, $\triangle\phi=|\phi_e - \phi_i|$, is larger than Planck mass. According to the Lyth bound, for such a model the tensor-to-scalar ratio should be bigger than $0.01$ \cite{Baumann}, that will be shown it is. On the other hand, the equation of state parameter, $\omega$, is very close to $-1$ at the beginning of inflation describing a quasi-de Sitter case, where the Hubble parameter, $H$, is thought as an almost constant parameter.  \\

\subsection{Inflationary Era and Observational Data}
Since the general form of the equations and perturbation parameters were acquired in the previous section,
we are going straight to the result of the proposed model. Inflation ends when the slow rolling parameter $\epsilon_H$ approaches unity.
To overcome the problem of standard cosmology models, the number of e-fold should be about $55-65$. Utilizing the definition of number of e-folds (\ref{efold}),
the initial scalar field could be extracted in term of $N$. The importance of the initial scalar field comes from the assumption
that the the quantum fluctuation is generated in the initial times of the inflation.

\begin{figure}[h]
\centering
\subfigure[ $r-n_s$ ]{\includegraphics[width=4.5cm]{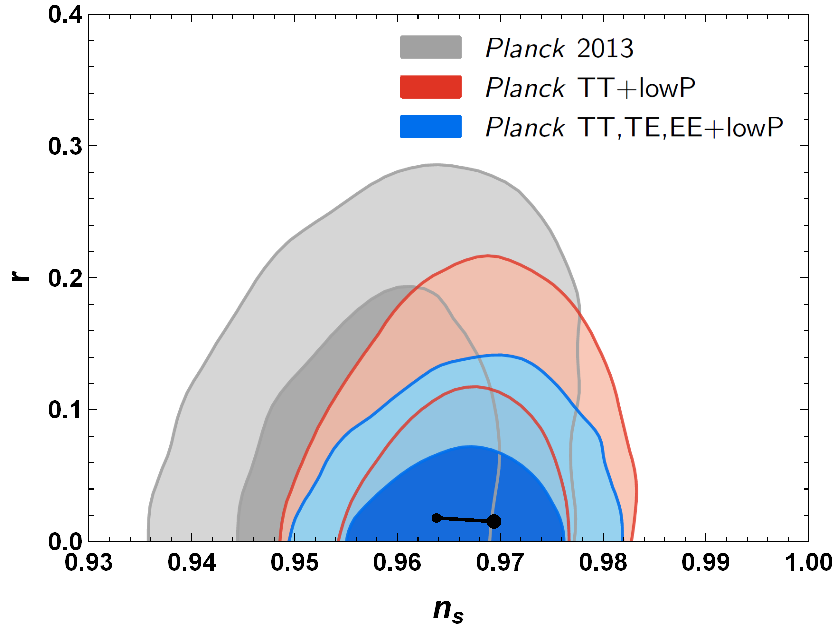}\label{rns}}
\hspace*{5mm}
\subfigure[ $dn_s/dlnk$ ]{\includegraphics[width=4.5cm]{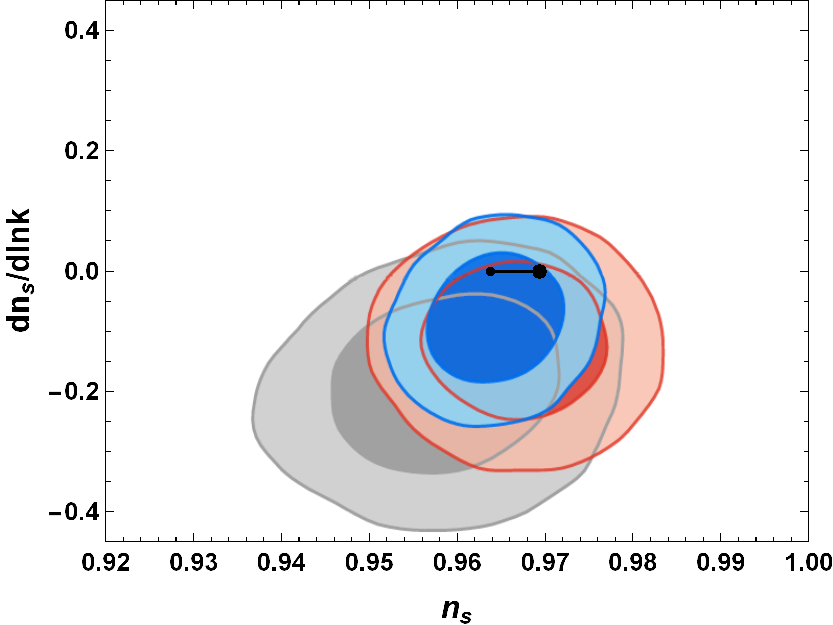}\label{dnsns}}
\caption{{\footnotesize a) $r$ b) ${dn_s \over dlnk}$ versus scalar spectra index $n_s$ have been depicted $N=55$(medium point) to $N=65$(large point), and the constant parameters: $\beta=9\times10^{-4}$, $\beta \mathcal{H}_1^2=1\times10^{-14}$.}}\label{rdnsns}
\end{figure}

Latest observational data brings a good insight about the evolution in one of its earliest time evolution.
From Planck data, the amplitude of scalar perturbation is about $\ln\Big(10^{10}\mathcal{P}^2_s \Big) = 3.094 \pm 0.034$ (Planck TT,TE,EE+lowP),
and the scalar spectra index, which is equal to one for a scale invariant spectrum, is measured about $n_s = 0.9645
\pm 0.0049$ ($68\%$ CL, Planck TT,TE,EE+lowP) \cite{planck2015}. In contrast with scalar perturbation,
Planck does not give an exact value for tensor-to-scalar ratio $r$;
it just specifies an upper bound for this parameter as $r < 0.10$ ($95\%$ CL, Planck TT,TE,EE+lowP) \cite{planck2015}. To test validity of the model as a candidate explaining the inflation, the perturbation parameters predicted by the model should be compared with the observational data. Amongst various data, the most important one is $r-n_s$ diagram in which the inflationary models could be organized based on their prediction about the diagram. Doing some manipulation, and using Eq.(\ref{ns}) and (\ref{r}), the parameters $r$ and $n_s$ could be estimated, in which the result has been prepared in Fig.\ref{rns}.\\

It could be realized that the model prediction about $r-n_s$ diagram is in perfect agreement with the Planck data, so that the results stands in $68\%$ CL area. In addition, the running scalar spectra index could be obtained using the same approach. Fig.\ref{dnsns} illustrates the result about the running of $n_s$ stays in the $68\%$ CL area of Planck data, implying another great consistency of the model with data. \\
The model could be checked one more time by comparison its estimation about amplitude of scalar perturbation with the Planck data. It the two following table the initial and final values of scalar field, tensor and scalar spectra indices, amplitude of scalar perturbation and tensor-to-scalar ratio have been presented for different values of the model free parameters.\\

\begin{table}[h]
  \centering
  {\footnotesize
  \begin{tabular}{p{1.6cm}||ccc}
    \toprule[1.5pt] \\[-3mm]
    $\beta \mathcal{H}_1^2$    &  $5\times 10^{-15}$    &   $7\times 10^{-15}$    &    $9\times 10^{-15}$   \\
      \midrule[1.5pt] \\[-3mm]
    $\phi_i/M_p$    &  $3.0971$      &    $3.1053$     &    $3.0939$     \\[0.5mm]
    $\phi_e/M_p$    &  $0.2980$      &    $0.2518$     &    $0.2221$     \\[0.5mm]
    $n_s$         &  $0.9668$      &    $0.9669$     &    $0.9667$     \\[0.5mm]
    $r$           &  $0.0331$      &    $0.0330$     &    $0.0332$     \\[0.5mm]
    $\ln\Big(10^{10}\mathcal{P}_s\Big)$     &  $3.6056$     &    $3.9520$       &    $4.1893$     \\[0.5mm]
    $n_T$         &  $-0.0165$     &    $-0.0165$    &    $-0.0166$    \\[0.1mm]
    \bottomrule[1.5pt]
  \end{tabular}
  }
  \caption{{\footnotesize The model prediction about initial and final values of scalar field, tensor and scalar spectra indices, amplitude of scalar perturbation and tensor-to-scalar ratio for constant parameters: $\beta=5\times10^{-4}$, $N=60$, and three different values of $\mathcal{H}_1$.}}\label{Table01}
\end{table}
\begin{table}[h]
  \centering
  {\footnotesize
  \begin{tabular}{p{1.6cm}||ccc}
    \toprule[1.5pt] \\[-3mm]
    $\beta$         &  $5\times 10^{-4}$    &   $7\times 10^{-4}$    &    $9\times 10^{-4}$   \\
      \midrule[1.5pt] \\[-3mm]
    $\phi_i/M_p$    &  $3.1098$      &    $3.0962$     &    $3.0886$     \\[0.5mm]
    $\phi_e/M_p$    &  $0.3141$      &    $0.2655$     &    $0.2341$     \\[0.5mm]
    $n_s$         &  $0.9670$      &    $0.9667$     &    $0.9666$     \\[0.5mm]
    $r$           &  $0.0329$      &    $0.0332$     &    $0.0333$     \\[0.5mm]
    $\ln\Big(10^{10}\mathcal{P}_s\Big)$     &  $3.5160$     &    $3.4991$       &    $3.4892$     \\[0.5mm]
    $n_T$         &  $-0.0164$     &    $-0.0166$    &    $-0.0166$    \\[0.1mm]
    \bottomrule[1.5pt]
  \end{tabular}
  }
  \caption{{\footnotesize The model prediction about initial and final values of scalar field, tensor and scalar spectra indices, amplitude of scalar perturbation and tensor-to-scalar ratio for constant parameters: $\mathcal{H}_1=3\times10^{-6}$, $N=60$, and three different values of $\beta$.}}\label{Table02}
\end{table}

It is clearly seen that all perturbation parameters predicted by the model for suggested free parameters are in the Planck data range, that in turn indicates the validity of the model as a suitable candidate for describing inflation as an earliest universe evolution.

\subsection{Attractor behavior}
The attractor behavior of the model has been studied in the earlier section generally. To specifically discuss about the feature, the introduced function of the Hubble parameter should be plugged in Eq.(\ref{attractor}), and show that the integral goes toward larger negative value by passing time. It is concluded that the integral ensures larger negative value by increasing time, which in turn exhibits the parameter $\delta H$ approaches zero with increasing time, and the attractor behavior could be satisfied by the model.  \\

\section{Conclusion}\label{t}
In this paper, we have analyzed the effect of having a minimum
measurable length scale on inflationary cosmology using inflationary cosmology in the Jacobian approach.
This approach is motivated by the relation between gravity and thermodynamics where Einstein field equations are  the equation of state for the geometry of spacetime. So, in this paper, the Friedmann equations are viewed as the  Clausius
relation. Then we have analyzed the modification of the Friedmann equations because of the existence of a minimum measurable
length. This consideration has been done by  analyzing  the effect of the existence of a minimum measurable length on the entropy
of the cosmological Horizon. Finally, these modified Friedmann equations are used for calculating the modifications to the
inflationary cosmology. This analysis is done using the Hamiltonian-Jacobi approach. We thus explicitly
calculate  the effect of having a minimum measurable length scale on inflationary cosmology.
It may be noted that property of the spectrum of large-scale magnetic fields which is generated due to the breaking of the
conformal invariance has been studied  in the context of inflationary cosmology \cite{z}. In this analysis, it has been demonstrated that the spectrum of the magnetic fields should not be perfectly scale-invariant. In fact, it was observed that this
spectrum should  be slightly red so that the amplitude of large-scale magnetic fields can be stronger
than a certain value. In this analysis, it was assumed that the absence of amplification occurs due to the late-time action of
some dynamo  mechanism. It would be interesting to repeat this analysis by assuming the existence
of a minimum measurable length scale. The  inflation has also been studied using a non-canonical
Lagrangian \cite{z1}-\cite{1z}. In this case, the modification to the kinetic term is modified.
This modification to the kinetic term depends only on the fields and not the derivatives of the fields. The non-canonical inflation has also been studied in the context of string-inspired inflation models \cite{za}.
The DBI action has also been used to study such models of inflation \cite{az}-\cite{a0}.
It has been demonstrated that the standard Hubble slow roll expansion to the non-canonical case can be generalized \cite{z2}. This generalization corresponds to the derivation of the  expressions for observables in terms of
the generalized slow roll parameters.  It would be interesting to analyze the models with
non-canonical kinetic term in the thermodynamic approach and also analyze the modification to
these models that can occur because of the existence of a minimum measurable length scale.\\
Utilizing Hamilton-Jacobi formalism in the introduced generalized Friedmann equation allows one to properly study inflation epoch of the Universe evolution. The calculated result demonstrates that the model could suitably describe inflation, and it sounds that the result is in consistence with the latest observational data of Planck. Considering the predicted potential behavior exhibits that the proposed model stands in \textit{Large Field Class} of inflationary models, so that the scalar field leaves the top of the potential at the beginning of inflation and slowly rolls down to the minimum. In addition, behavior of equation of state parameter $\omega$ is in agreement with whole work. It expresses a quasi-de Sitter expansion for the universe, in which at the initial of inflation, the parameter is so close to $-1$.  The tensor-to-scalar ratio and the running scalar spectra index were plotted versus $n_s$. For the both, the result stands in the $68\%$ CL area, a great consistency of the model prediction with the Planck data. The main output of the model about inflationary era were briefly prepared in Tables.\ref{Table01} and \ref{Table02} for different values of the free parameters. The estimated values for $n_s$ and $\ln\Big(10^{10}\mathcal{P}_s\Big)$ indicate that these parameters are respectively almost about $n_s=0.9667$ and $3.605$; which are in good agreement with Planck data, where $n_s=0.9645 \pm 0.0049$ and $\ln\Big(10^{10}\mathcal{P}_s\Big)= 3.094 \pm 0.034$. Finally, investigation of attractor behavior for the model comes to another pleasant outcome. The homogenous perturbation $\delta H$ approaches zero with increasing time for all three cases describing that the model could satisfy the attractor behavior.\\

\subsection*{Acknowledgments}
The research of AFA is supported by Benha University(www.bu.edu.eg). The authors would like to thank the referee for
the useful suggestions which helped to improve the paper.













\end{document}